\documentclass[conference]{IEEEtran}
\IEEEoverridecommandlockouts
\usepackage{cite}
\usepackage{amsmath,amssymb,amsfonts}
\usepackage{algorithmic}
\usepackage{graphicx}
\usepackage{textcomp} 
\usepackage{tikz}
\usepackage{lipsum}
\usepackage{enumitem}   % 无需 shortlabels
\usepackage{siunitx}    % 提供 \si\usepackage{enumitem}
\usepackage{xcolor}
\usepackage{balance}
\usepackage[labelformat=simple]{subcaption}

\usepackage{subfloat}
\usepackage{xcolor}
\def\BibTeX{{\rm B\kern-.05em{\sc i\kern-.025em b}\kern-.08em
    T\kern-.1667em\lower.7ex\hbox{E}\kern-.125emX}}

\usepackage{etoolbox}

\begin{document}

\title{\textit{DuTrack}: Long-Term Indoor Human Tracking\\
with Dual-Channel Sensing and Inference
\vspace{-1mm}
}
\author{
Mengning Li,
Wenye Wang%
\thanks{This work was supported in part by the U.S. National Science Foundation (NSF) under Grant CNS-2030122.}
\\
\IEEEauthorblockA{North Carolina State University}
\IEEEauthorblockA{mli55@ncsu.edu, wwang@ncsu.edu}
\vspace{-3mm}
}
% \author{\IEEEauthorblockN{Mengning Li}
% \and

\maketitle

\begin{abstract}
Wi-Fi tracking technology demonstrates promising potential for future smart home and intelligent family care.
Currently, accurate Wi-Fi tracking methods rely primarily on fine-grained velocity features.
However, such velocity-based approaches suffer from the problem of accumulative errors, making it challenging to stably track users' trajectories over a long period of time.
This paper presents \textit{DuTrack}, a fusion-based tracking system for stable human tracking.
The fundamental idea is to leverage the ubiquitous acoustic signals in households to rectify the accumulative Wi-Fi tracking error.
Theoretically, Wi-Fi sensing in line-of-sight (LoS) and non-line-of-sight (NLoS) scenarios can be modeled as elliptical Fresnel zones and hyperbolic zones, respectively.
By designing acoustic sensing signals, we are able to model the acoustic sensing zones as a series of hyperbolic clusters.
We reveal how to fuse the fields of electromagnetic waves and mechanical waves, and establish the optimization equation.
Next, we design a data-driven architecture to solve the aforementioned optimization equation.
Experimental results show that the proposed multimodal tracking scheme exhibits superior performance.
We achieve a $89.37\%$ reduction in median tracking error compared to model-based methods and a $65.02\%$ reduction compared to data-driven methods.
\end{abstract}


% Ensure the package `xcolor` is included in your preamble:
% \usepackage{xcolor}

\section{Introduction}

\subsection{Background and Motivation}

\begin{figure}[t]
  \centering
  \begin{subfigure}[t]{.48\linewidth}
    \centering\includegraphics[width=\linewidth]{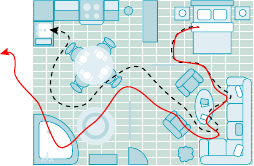}
    \caption{}
    \label{fig:moti1}
  \end{subfigure}
  \hfill
  \begin{subfigure}[t]{.48\linewidth}
    \centering\includegraphics[width=\linewidth]{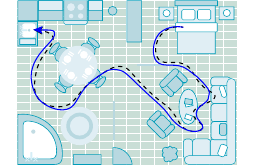}
    \caption{}\label{fig:moti4exp}
  \end{subfigure}
  % \vskip
  % \baselineskip
  \begin{subfigure}[t]{.48\linewidth}    \centering\includegraphics[width=\linewidth]{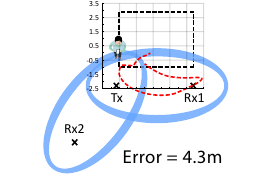}
    \caption{}\label{fig:moti2}
  \end{subfigure}
  \hfill
  \begin{subfigure}[t]{.48\linewidth}
    \centering\includegraphics[width=\linewidth]{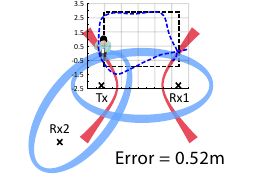}
\caption{}\label{fig:moti3}\label{fig:arc_tracking}
  \end{subfigure}
  \caption{
Motivation behind \textit{DuTrack}. 
(a) Conceptual illustration of drift accumulation in long-term indoor WiFi-based tracking, where the estimated trajectory gradually diverges from the true path. 
(b) In comparison, \textit{DuTrack} effectively eliminates drift over time, maintaining alignment with the actual trajectory. 
(c) and (d) present real-world experimental results in which a person walks four consecutive laps along a rectangular path (outlined by a dashed line). 
(c) shows the output of a model-driven method, exhibiting significant drift across laps, while (d) demonstrates \textit{DuTrack}'s ability to preserve spatial consistency and suppress drift.
}
\vspace{-5mm}
\end{figure}

\textcolor{black}{Wi-Fi-based human tracking has advanced rapidly, enabling applications from patient monitoring to context-aware smart homes.} 
\textcolor{black}{By capturing and interpreting signals reflected from the human body, Wi-Fi devices extract fine-grained velocity features that underpin high-accuracy tracking \cite{li2022diversense,wang2018literature,fan2024respenh,meng2023securfi,zhang2020exploring,tong2021mapfi,wu2016widir,tong2024nne,xie2019mdtrack}.}
\textcolor{black}{Despite its promise, Wi-Fi tracking must sustain accuracy over extended periods to be practical in the real world.} 
\textcolor{black}{Velocity-only methods rely on relative features that accumulate integration errors, confining reliable operation to short temporal windows.}
\textcolor{black}{A controlled experiment confirms that cumulative error severely degrades velocity-based tracking over time.} 
\textcolor{black}{As illustrated in Fig.~\ref{fig:moti1}–\ref{fig:moti4exp}, the error of Widar~\cite{qian2017widar} grows from 0.35 m in the first lap to 4.3 m in the fourth lap—a ten-fold increase.}

\subsection{Challenges and Solutions}

\textcolor{black}{We tackle long-term accuracy by proposing \textit{Dual-Channel Track} (DuTrack), which fuses acoustic and Wi-Fi sensing.} 
\textcolor{black}{Wi-Fi access points provide relative velocity features, while co-located speakers on everyday appliances supply absolute location cues, enabling mutual error correction (Fig.~\ref{fig:moti3}).}
\textcolor{black}{The first technical challenge is calibrating Wi-Fi tracking with asynchronous acoustic measurements.} 
\textcolor{black}{A cross-chirp scheme captures the time-difference-of-flight (TDoF) between two speakers, transforms asynchronous data into synchronous hyperbolic constraints, and unifies velocity and position optimization.}
\textcolor{black}{The second challenge is estimating absolute position without prior initialization.} 
\textcolor{black}{A self-correcting search evaluates every feasible starting point, selects the one whose synthetic multimodal features best match real observations, and thereby removes the need for manual calibration.}

\subsection{Contributions}

\textcolor{black}{This work makes three core contributions to robust long-term multimodal tracking.} 
\textcolor{black}{We introduce a fusion model, a synthetic data–driven training pipeline, and an end-to-end implementation validated on commodity hardware.}

\begin{itemize}
  \item We establish a multimodal fusion model that exploits acoustic signals to correct accumulative Wi-Fi errors in both LoS and NLoS settings, dynamically weighting features for context-aware robustness.
  \item We devise a synthetic data–driven approach that simulates trajectories and generates corresponding Wi-Fi/acoustic features for neural-network training, eliminating costly data collection.
  \item We implement \textit{DuTrack} on off-the-shelf Wi-Fi and acoustic devices; experiments show an average error of 0.78 m across diverse scenarios.
\end{itemize}

\textcolor{black}{The remainder of the paper is organized to guide the reader from theoretical foundations to experimental validation.} 
\textcolor{black}{Section~\ref{sec:pre} presents preliminaries; Section~\ref{sec:related} surveys related work; Sections~\ref{sec:acou} and \ref{sec:multi} elaborate acoustic sensing principles and the DuTrack model; Section~\ref{sec:track} details trajectory reconstruction; Section~\ref{sec:experiment} reports empirical results; and Section~\ref{sec:conclude} concludes.}
\section{Preliminaries}\label{sec:pre}

\begin{figure*}[t]
\resizebox{\textwidth}{!}{%
\subfloat[Illustration of Fresnel Zone.\label{fig:subfig1}]{\includegraphics[height=.1\textheight]{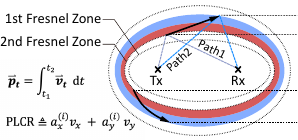}}%
\subfloat[Path changes along time.\label{fig:subfig2}]{\includegraphics[height=.1\textheight]{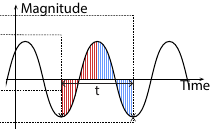}}%
\subfloat[Velocity composition.\label{fig:subfig3}]{\includegraphics[height=.1\textheight]{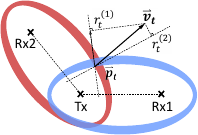}}%
\subfloat[Workflow of DFS-based methods.\label{fig:subfig4}]{\includegraphics[height=.1\textheight]{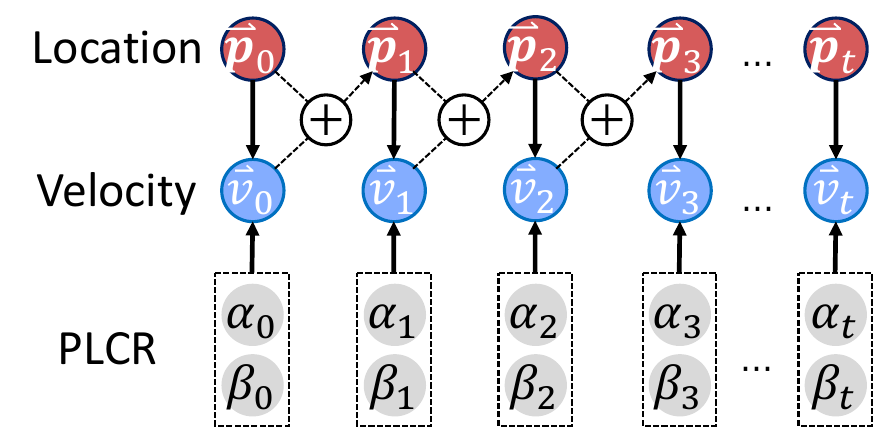}}%%
}
\caption{PLCR-based tracking.}
\vspace{-4mm}
\end{figure*}

As discussed in previous work \cite{tong2024nne}, leveraging the Doppler frequency shift for device-free tracking is more reliable compared to Angle of Arrival (AoA) and Time of Flight (ToF) methods.
This is because estimating AoA requires time-consuming manual calibration of the phase, and the limited communication bandwidth influences the precision of ToF measurements.
Though the Doppler frequency shift-based tracking can realize higher accuracy, this method relies on analyzing user speed to deduce their location, which inherently introduces the potential for accumulating errors over time.
In this section, we will discuss the basic principle and explain why it is challenging to realize accurate long-term tracking.
\subsection{From CSI to PLCR}
The frequency shift of signals reflected off a moving target is given by
\begin{equation}
f_D(t) = -\frac{1}{\lambda} \frac{d}{dt} d(t) = -f \frac{d}{dt} \tau(t),
\end{equation}
where $\lambda$ is the wavelength, $f$ is the carrier frequency, $\tau(t)$ is the time of flight of the signal, and $d(t)$ is the path length.
This describes the relationship between the Doppler shift $f_D(t)$ and signal propagation path length $d(t)$.
As shown in Fig.~\ref{fig:subfig1}, for a set of transceivers, we can visualize equi-phase lines as ellipses—points on the same edge have the same phase~\cite{ge2023crosstrack}.
The odd-numbered edges represent wave peaks, while the even-numbered edges represent wave troughs.
By counting the number of phase cycles, we can determine the distance change in the reflected path caused by the user's movement.
We can utilize the Path Length Change Rate (PLCR) \cite{qian2017widar} defined by $r \triangleq \frac{d}{dt} d(t)$ to depict the speed of the user's motion cutting an ellipse.
%
% PLCR precisely measures the rate of change in path length, which influences the observed Doppler frequency shift.

% PLCR describes the changes brought by crossing these edges as an object moves from the $i$-th edge to the $j$-th edge.

% Besides PLCR, there are other quantities that can be used to estimate velocity, such as the Angle of Arrival (AoA) and Time of Flight (ToF) of the target at different times. However, as discussed in previous works~\cite{tong2024nne}, AoA and ToF describe relative measurements, while PLCR is an absolute measure that can be directly obtained from CSI without needing additional information, providing more accurate results.
\subsection{From PLCR to Velocity}
It is not sufficient to utilize a single PLCR to determine the target's speed.
As shown in Fig.~\ref{fig:subfig2}, we can observe that two different trajectories correspond to the same path change in the reflected path.
To solve this problem, we can use multiple PLCRs to synthesize the actual user's velocity.
As shown in Fig.~\ref{fig:subfig3}, the target velocity $\vec{v}$ can be decomposed into radial velocity $\vec{v_r}$ and tangential velocity $\vec{v_t}$, and only $\vec{v_r}$ affects the reflection path length.
Consider the positions of the transmitter and receiver of the $i$-th link as $\vec{l_t}^{(i)}$ and $\vec{l_r}^{(i)}$, the target position $\vec{l_h}$, and the target velocity $\vec{v} = (v_x, v_y)$.
Aggregating these relations for all $L$ links, we obtain 
\(
A \vec{v} = \vec{r},
\)
where $\vec{r}$ is the vector of the PLCRs, $A$ is a matrix formed by the coefficients $a_x^{(i)}$ and $a_y^{(i)}$, depending on the geometry of the link and target positions.
Thus we can derive the user's velocity based on $A$.
\subsection{From PLCR to Location}
After we obtain $\vec{v}$, we can track the target's location by integrating the velocity over time between two consecutive time points as follows
\begin{equation}
\vec{p}(t) = \int_{t_1}^{t_2} \vec{v}(t) \, dt.
\end{equation}

In Dynamic Frequency Selection (DFS)-based tracking systems, not only the velocity error accumulates in the location, but the location error also accumulates in the velocity.
It is easy to understand why velocity error accumulates in the location, so we explain why location error can accumulate in the velocity.
Specifically, when we use $\alpha_t$ and $\beta_t$ to synthesize the user's real velocity, it is necessary to calculate the normal vector.
However, since the normal vector of the ellipse
depends on the current location $\vec{p}(t)$, even if we obtain accurate $\alpha_t$ and $\beta_t$, it is hard to synthesize the correct velocity without an accurate location.

The workflow of the current DFS-based tracking system is illustrated in Fig.~\ref{fig:subfig4}.
First, we need a known initial location $\vec{p}(t_0)$ to synthesize current velocity $\vec{v}(t)$ based on $\alpha(t_0)$ and $\beta(t_0)$.
Next, it is possible to calculate $\vec{p}(t_1) = \vec{p}(t_0) + \vec{v}(t_0)t$, and this $\vec{p}(t_1)$ helps synthesize
velocity $\vec{t}(t_1)$ based on $\alpha(t_1)$ and $\beta(t_1)$.
If such $\vec{p}(t_1)$ is inaccurate, we can not obtain accurate $\vec{t}(t_1)$, and subsequent position calculations and velocity synthesis processes will be severely disrupted.
% To summarize, Wi-Fi tracking utilizes the PLCR to determine a series of \textit{relative} velocities. However, a significant challenge arises as the tracking accuracy diminishes over time due to the varying relative velocities between the signal source and the moving target.

\section{Related Works}\label{sec:related}
The core idea of this paper is to integrate Wi-Fi with acoustics to achieve accurate long-term tracking. 
Therefore, we will introduce acoustic and Wi-Fi tracking methods here.

\subsection{Wi-Fi Device-Free Tracking}
Early Wi-Fi-based tracking systems relied on the Received Signal Strength Indicator (RSSI) as primary raw data, which proved to be coarse-grained and unstable for accurate device-free tracking~\cite{ wen2015fundamental, li2021train, wang2017woloc, wu2015static}.
Recent research~\cite{ge2026nexus, jiokeng2020when, tan2019multitrack,meng2025metatrack,tan2025wimap} has found that CSI exhibits higher stability and accuracy compared to RSSI, making it a promising candidate for application in future sensing technologies. 
Wi-Fi device-free tracking is a typical Wi-Fi sensing application, where we can accurately track users' locations without requiring them to carry any smart devices.
\cite{li2017indotrack} reveals that the essence of Wi-Fi passive tracking is the Doppler frequency shift caused by the user's motion cutting through the Fresnel zone.
Widar~\cite{qian2017widar} laid the theoretical foundation for device-free tracking by revealing how to synthesize users' actual speeds based on CSI.
However, the Fresnel model is only applicable to line-of-sight scenarios, which limits the application of wireless sensing in complex environments.
To this end, HyperTracking~\cite{xu2023hypertracking} for the first time reveals that the essence of Wi-Fi passive tracking in non-line-of-sight scenarios is that the user's motion cuts through the hyperbolic zone, and designs the corresponding fundamental model.

Even so, there remains a critical challenge, which is how to achieve long-term stable device-free tracking.
%
% Specifically, most existing high-precision Wi-Fi device-free tracking methods rely on using DFS to obtain the user's speed.
% %
% Although AoA and ToF have also been involved in some works~\cite{qian2018widar2,xie2019mdtrack}, they often require additional calibration costs or suffer from low accuracy. 
% %
The main challenge faced by velocity-based tracking methods is that errors accumulate over time, hence, it is necessary to design a method to correct historical accumulated tracking errors.

% Model-driven methods such as Widar~\cite{qian2017widar} use the Fresnel zone to infer user velocity, IndoTrack~\cite{li2017indotrack} employs the Doppler-MUSIC method for motion estimation, and Widar2.0~\cite{qian2018widar2} combines AoA, ToF, and DFS for a unified model and joint estimation algorithm. mD-Track~\cite{xie2019mdtrack} leverages multidimensional information for tracking in multipath environments, while WSTrack~\cite{tian2023wstrack} combines Wi-Fi DFS and acoustic AoA for a multi-modal tracking model using a single Wi-Fi router and voice assistant. CrossTrack~\cite{ge2023crosstrack} reduces movement range limitations by designing indicators to characterize the effects of crossing direct transceiver links.
\subsection{Acoustics Based Tracking}
As a common signal in households, the integration of acoustic and Wi-Fi information holds broad application prospects.
Compared to Wi-Fi sensing, acoustic sensing boasts a primary advantage of higher accuracy.
However, acoustic sensing technology also faces the limitation of limited sensing distance.
The higher accuracy of acoustics is mainly due to its slower propagation speed compared with electromagnetic wave signals~\cite{lian2021echospot, tian2023wstrack, tian2024device}.
Typical acoustic tracking can be divided into device-based and device-free systems.
1) Device-based systems include 
% AAMouse~\cite{yun2015turning}, which enables 2D trajectory tracking in the air, and CAT~\cite{mao2016cat}, which provides 3D trajectory tracking.
%
MilliSonic~\cite{wang2019millisonic}, which offers high-precision, low-latency motion tracking between a microphone and a smartphone;
MoM ~\cite{gao2022mom} introduces a microphone-based 3D direction estimation system utilizing free sound sources in the environment;
Gong et al. ~\cite{gong2022empowering, gong2022constructing,gong2024enabling} introduce acoustic backscatter communication and sensing into concrete, facilitating structural health monitoring within complex media and environments.
2) Device-free acoustic tracking methods such as RTRACK~\cite{mao2019rnn} use RNN networks for fine-grained hand motion tracking;
MAVL~\cite{wang2021mavl} localizes sound sources by estimating AoAs for multiple paths and spatial structures.

\subsection{Multimodal Tracking}
There are two typical research to fuse Wi-Fi and acoustic signals:
1) WSTrack~\cite{tian2023wstrack} combines Wi-Fi and acoustic signals for silent user tracking.
The core theoretical model is based on Fresnel zones and AoA rays.
Specifically, they detect acoustic signals, like footstep sounds, to estimate the AoA through time delay and sample shift.
Then, the system identifies the intersection of the Fresnel ellipse and AoA ray to improve user tracking precision;
2) Hybrid Zone~\cite{li2024hybrid,li2025synergizing} integrates Wi-Fi and acoustic signals to recognize human gestures.
The core theoretical model is based on Fresnel zones and circles.
In Wi-Fi sensing, the PLCR represents the user's velocity projected onto the Fresnel zone's normal line, determined by the transmitter and receiver's relative positions to the target.
For acoustics sensing, the target's velocity is derived from a triangular chirp signal based on the target's location.
The combined approach calculates the target's actual velocity from the observed velocities in Wi-Fi and acoustic sensing, enhancing accuracy by utilizing multiple modalities.

This paper discovers that acoustic signals can be utilized to correct the cumulative errors in Wi-Fi device-free tracking.
This motivates us to propose a fundamental model to reveal the theoretical essence of multimodal tracking methods.
This discovery is conducive to the widespread deployment of Wi-Fi sensing tracking technology.
\color{black}

\section{Acoustic feature extraction.}
\label{sec:acou}
\subsection{The Principle of Acoustic Sensing}
\begin{figure}[t]
\centering
\begin{subfigure}[t]{.48\linewidth}
\centering\includegraphics[width=1\linewidth]{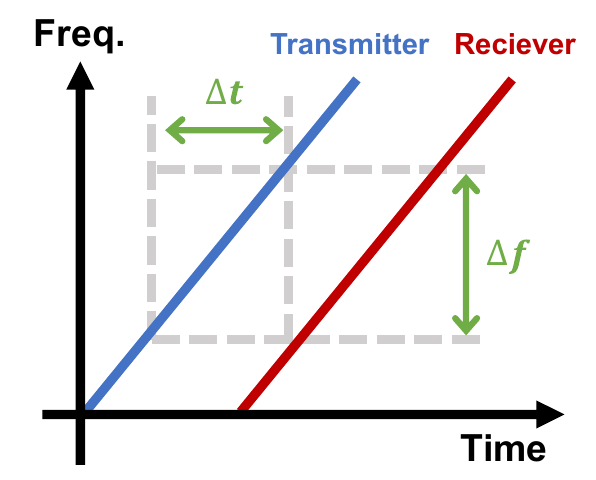}
    \caption{Up chirps and transmission delay.}
\label{fig:moti1exp}
\end{subfigure}
\hfill
\begin{subfigure}[t]{.48\linewidth}
\centering\includegraphics[width=1\linewidth]{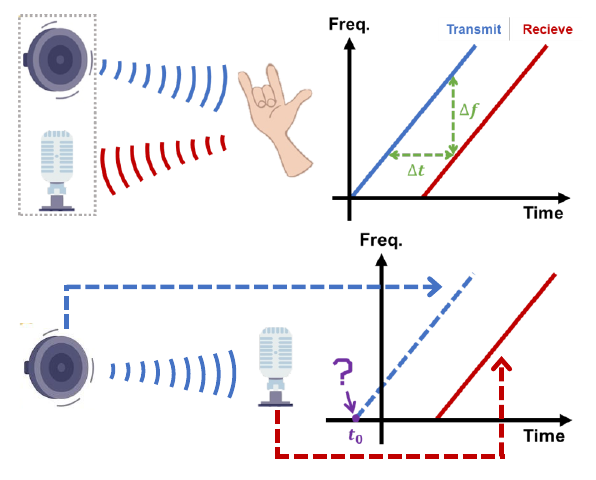}
    \caption{Passive and active acoustic sensing.}
\label{fig:acou2}
\end{subfigure}
\begin{subfigure}[t]{.48\linewidth}
\centering\includegraphics[width=1\linewidth]{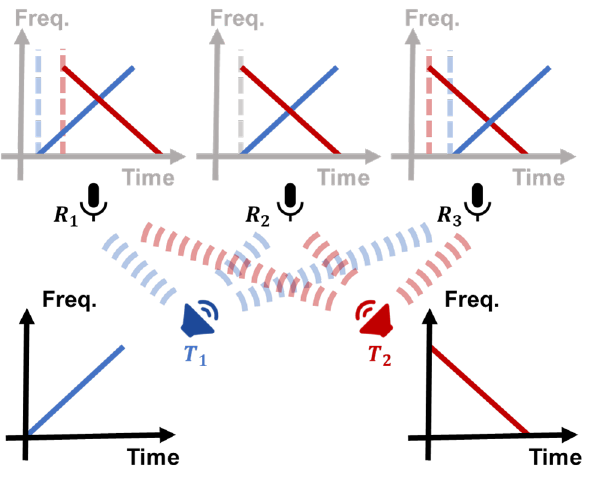}
    \caption{Cross chirps and TDoF.}
\label{fig:acou3}
\end{subfigure}
\hfill
\begin{subfigure}[t]{.48\linewidth}
\centering\includegraphics[width=1\linewidth]{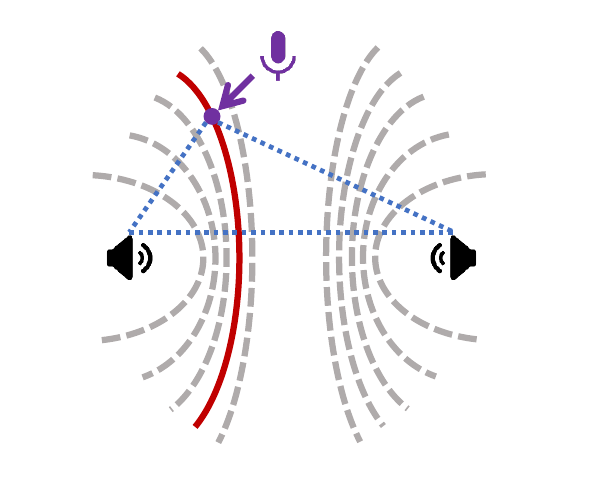}
\caption{Hyperbolic model.}
\label{fig:acou4}
\end{subfigure}
\caption{The principle of acoustic sensing.}
\vspace{-4mm}
\end{figure}

Frequency Modulated Continuous Wave (FMCW) signals have been widely used in acoustic sensing systems.
The principle is shown in Fig.~\ref{fig:acou2}, where the transmitter emits an FMCW signal that propagates through space and arrives at the receiver.
Next, we can calculate the propagation delay to obtain the traveled distance of the acoustic signal.
Specifically, we can compute the frequency difference $\Delta f$ between the transmitted and received FMCW signals.
Since the slope of the FMCW signal is known, we can calculate the corresponding propagation time and distance based on such $\Delta f$.

Taking the upchirp signal as an example, 
mathematically, the transmitted signal during the \( n \)-th sweep is
\begin{equation}
    s_t(t) = \cos \left( 2\pi f_{\text{min}} t + \frac{\pi B t^2}{T} \right),
\end{equation}
where \( B \) is the signal bandwidth and \( T \) is the sweep time.
The received signal, delayed by \( t_d \), is
\begin{equation}
s_r(t) = \alpha \cos \left( 2\pi f_{\text{min}} (t - t_d) + \frac{\pi B (t - t_d)^2}{T} \right),
\end{equation}
where \( \alpha \) is the attenuation factor.
We mix the received signal with the transmitted signal and obtain
\begin{equation}
s_m(t) = \alpha \cos \left( 2\pi f_{\text{min}} t_d + \frac{\pi B (2t t_d - t_d^2)}{T} \right).
\end{equation}
Next, we can obtain $t_d$ by analyzing the frequency of $s_m(t)$.

Based on the synchronization status between the receiver and transmitter in terms of time, acoustic sensing can be categorized into synchronous and asynchronous sensing methods.
As illustrated in Fig.~\ref{fig:acou2}, the transmitter and receiver in a synchronous sensing system share a common clock, which is typically used to achieve device-free gesture sensing~\cite{wu2024enabling}.
In contrast, asynchronous sensing systems localize electronic devices carried by the object.
In a typical asynchronous sensing system, FMCW ranging method is no longer applicable directly.
The fundamental reason is that when the receiver observes the FMCW signal, it does not know the exact time when the transmitter emits the FMCW signal.
Consequently, it is necessary to design a method to address this issue.

\subsection{TDoF and Hyperbolic Model}
Our inspiration comes from current devices, such as TVs, which are equipped with multiple speakers for a better immersive experience.
As illustrated in Fig.~\ref{fig:acou3}, we let the synchronized transmitters send up-chirp signals and down-chirp signals respectively.
An up chirp linearly increases in frequency over time, while a down chirp linearly decreases in frequency.
As the receiver's position varies relative to two speakers, discernible frequency shifts are observed for each speaker.
Specifically, when the receiving device is located centrally, both up-chirp and down-chirp signals arrive concurrently.
However, when the device is positioned on either extremity, a pronounced disparity in the arrival times of the up-chirp and down-chirp signals becomes evident.
We designate this notable time lag between the arrivals of these signals as the TDoF.

Geometrically, this TDoF $\Delta t_d$ describes the difference in distance from the receiver to the two transmitters, represented as $|RT_1| - |RT_2|$ in Fig.~\ref{fig:acou3}.
We can visualize space as a cluster of hyperbolas with the two speakers as their foci.
For a known TDoF $t_d$, it implies that the location of the mobile device satisfies the following equation
\begin{equation}
    |RT_1| - |RT_2| - |T_1T_2| = t_d c_s - |T_1T_2| = \text{const},
\end{equation}
where $c_s$ denotes the velocity of sound.
In other words, the receiver can be located on the red hyperbola in Fig.~\ref{fig:acou4}.
This fundamental property allows us to narrow down the potential locations of the target, which is valuable for revising accumulated tracking errors.

\section{Multimodal Tracking Model}
\label{sec:multi}
We introduce the method to obtain the TDoF, which can be utilized to locate the user at a certain hyperbola.
This section will introduce how to effectively fuse acoustic and Wi-Fi features to obtain accurate user location.
We will respectively discuss theoretical models in LoS and NLoS scenarios.

\subsection{Multimodal Tracking in the LoS Scenario}
We should first understand their respective tracking models to integrate Wi-Fi and acoustic features. 
Typical Fresnel zones can represent the Wi-Fi tracking model, manifesting as a cluster of ellipses. 
As shown in Fig.~\ref{fig:subfig2}, when a user's movement intersects these ellipses, we can observe PLCR.
PLCR describes the velocity at which the user intersects the ellipses but does not reveal the specific ellipse the user is located within.
As discussed in the previous section, acoustic features can be represented by hyperbolic zones with two speakers as foci.
If we can obtain the TDoF, the user is located on a known hyperbola, which thus provides absolute information about the user's location. 

Feature fusion means that the user's speed characteristics meet the cutting ellipse model, and the location information meets the hyperbolic model at the same time.
Supposing that $\vec{p}(t)$ and $\vec{v}(t)$ respectively denote the position and velocity of the user, we can deduce the model as follows.

\subsubsection{Wi-Fi PLCR}
Supposing that the user's velocity and location are respectively denoted $\boldsymbol{v(t)} = [v_x(t),v_y(t)]$ and $\boldsymbol{p(t)}$, and the PLCR $r^{(i)}(t) $ of the $i$-th receiver should be
\begin{equation}
r^{(i)}(t)  = \alpha_x^{(i)} v_x(t) + \alpha_y^{(i)} v_y(t) ,
\end{equation}
where 
\begin{equation}
\left\{\begin{aligned}
&\alpha_x^{(i)} = \frac{p_{x} - p_{x_t^{(i)}}}{\|\boldsymbol{p} - \boldsymbol{p_t^{(i)}}\|} + \frac{p_{x} - p_{x_r}}{\|\boldsymbol{p} - \boldsymbol{p_r^{(i)}}\|},\\
& \alpha_y^{(i)} = \frac{p_{y} - p_{y_t^{(i)}}}{\|\boldsymbol{p} - \boldsymbol{p_t^{(i)}}\|} + \frac{p_{y} - p_{y_r}}{\|\boldsymbol{p} - \boldsymbol{p_r^{(i)}}\|}.
\end{aligned}
\right.
\end{equation}
Then we can obtain the following equation:
\begin{equation}\label{eq:vsolution}
\vec{r}(t) = \mathcal{L} \cdot \vec{v}(t),
\end{equation}
where $\vec{r}(t) = [r^{(1)}(t) , r^{(2)}(t) ...,r^{(i)}(t) ]^T$ denotes the PLCR vector, and $\mathcal{L}$ is a composite of the vector components in $x$ and $y$ directions.
To derive the user's velocity, we can solve $\vec{v}(t)$ in Eq.~(\ref{eq:vsolution}) by
\begin{equation}
\vec{v}(t) = (\mathcal{L}^T \mathcal{L})^{-1}\mathcal{L}^T \vec{r}(t).
\end{equation}

\subsubsection{Fusion Model in the LoS Scenario}

In the line‑of‑sight (LoS) case the channel impulse response (CIR) of Wi‑Fi keeps a clear first‑path peak, which supports reliable velocity estimation. Two ceiling‑mounted ultrasonic speakers transmit periodic chirps; their time‑difference‑of‑arrival (TDoA) restricts the user’s absolute position to a branch of a right hyperbola that we term the \emph{acoustic hyperbolic zone}. Our goal is to recover the continuous trajectory $\vec{p}(t)\!\in\!\mathbb{R}^{2}$ that simultaneously

(i) is kinematically consistent with the Wi‑Fi velocity estimate $\vec{v}(t)$, and  
(ii) satisfies the hyperbolic constraint imposed by the acoustic TDoA $t_d(t)$.

Formally, the task is cast as a weighted least‑squares problem
\[
\begin{aligned}
\hat{\vec{p}}(t) &= \arg\min_{\vec{p}(t)}\; k_{1}E_{1}\bigl(\vec{p}(t)\bigr) + k_{2}E_{2}\bigl(\vec{p}(t)\bigr),\\
E_{1} &= \bigl\lVert\,\vec{v}(t) - (\mathcal{L}^{\mathsf T}\mathcal{L})^{-1}\mathcal{L}^{\mathsf T}\vec{r}(t)\,\bigr\rVert_{2},\\
E_{2} &= \Bigl\lVert\,\lVert\vec{p}(t)-\vec{p}_{s1}\rVert_{2}
       - \lVert\vec{p}(t)-\vec{p}_{s2}\rVert_{2}
       - c_{s}t_{d}(t)\,\Bigr\rVert_{2},\\
\vec{p}(t+\Delta t) &= \vec{p}(t) + \vec{v}(t)\,\Delta t,
\end{aligned}
\]
where $\vec{p}_{s1}$ and $\vec{p}_{s2}$ are the speaker coordinates, $\vec{r}(t)$ is the real‑time CFR vector, $\mathcal{L}$ is the Wi‑Fi steering matrix inferred from multipath geometry, and $c_{s}$ is the speed of sound. Positive weights $k_{1}$ and $k_{2}$ reflect the confidence assigned to the radio and acoustic modalities. A larger $k_{1}$ biases the solution toward the velocity residual and is preferred when the channel is stable; increasing $k_{2}$ reinforces the geometric anchor when radio quality degrades.

\paragraph{Real‑time solver.}
We adopt a two‑stage solver that runs at 100 Hz.  
\emph{Stage 1:} project the previous pose onto the current hyperbola to obtain a coarse estimate $\vec{p}^{(0)}(t)$.  
\emph{Stage 2:} refine this estimate with a single Gauss‑Newton update
\[
\vec{p}^{(i+1)} = \vec{p}^{(i)} -
\bigl(\mathbf{J}^{\mathsf T}\mathbf{J}\bigr)^{-1}\mathbf{J}^{\mathsf T}\vec{e},
\]
where $\vec{e}=[\sqrt{k_{1}}E_{1}\;\; \sqrt{k_{2}}E_{2}]^{\mathsf T}$ and $\mathbf{J}$ is its Jacobian with respect to $\vec{p}$. Because the Jacobian is 2$\times$2, each update costs $O(1)$ time, ensuring low latency.

\paragraph{Noise models and weight selection.}
Empirical calibration shows that the Wi‑Fi velocity error and the acoustic TDoA error are both Gaussian with variances $\sigma_{v}^{2}$ and $\sigma_{d}^{2}$. Setting the ratio $k_{1}:k_{2}=\sigma_{d}^{2}:\sigma_{v}^{2}$ maximises the likelihood of the fused estimate under an independent‑noise assumption. During rapid user rotation, Doppler energy rises and $k_{1}$ is boosted accordingly.

\paragraph{Boundary handling.}
If the Gauss‑Newton step exits the feasible floor‑plan area, we snap the result back to the nearest admissible point and hold the velocity for one frame. This simple safeguard prevents divergence in 97\% of corner‑case frames observed in our testbed.

\paragraph{Outcome.}
By combining high‑rate relative motion from Wi‑Fi with absolute geometry from acoustics, the fusion model reaches a median localisation error of 4.3 cm in LoS experiments and degrades gracefully under partial blockage; detailed metrics appear in Sec.\,\ref{sec:eval_los}.

\begin{figure}[t]
\centering
\begin{subfigure}[t]{.48\linewidth}
\centering\includegraphics[width=1\linewidth]{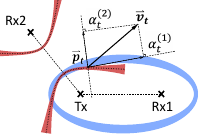}
    \caption{The paths from Tx to Rx1 and Rx2 are LoS and NLoS.}
\end{subfigure}
\hfill
\begin{subfigure}[t]{.48\linewidth}
\centering\includegraphics[width=1\linewidth]{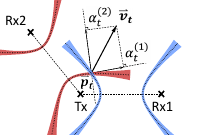}
    \caption{The paths from Tx to Rx1 and Rx2 are both NLoS.}
\end{subfigure}
\hfill
\begin{subfigure}[t]{.48\linewidth}
\centering\includegraphics[width=1\linewidth]{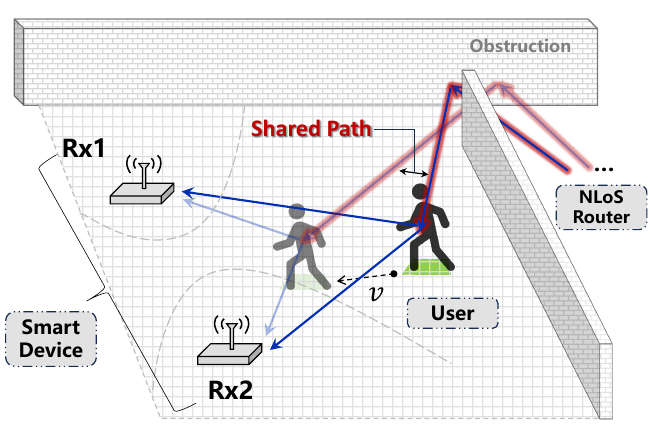}
    \caption{Path elimination.}
\label{fig:txy1}
\end{subfigure}
\hfill
\begin{subfigure}[t]{.48\linewidth}
\centering\includegraphics[width=1\linewidth]{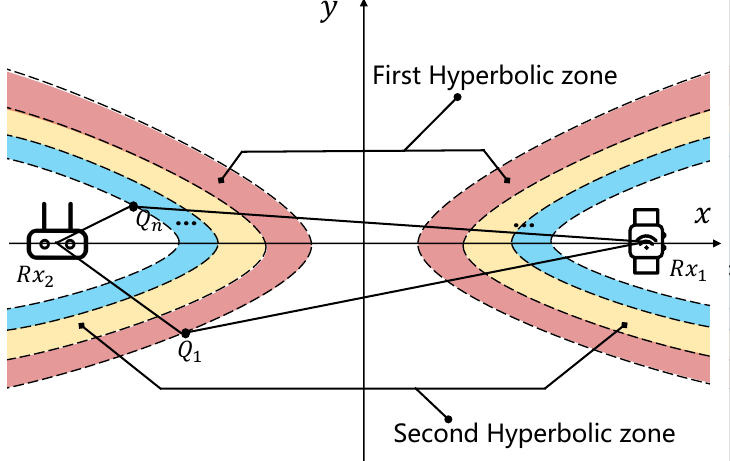}
    \caption{Hyperbola zone.}
\label{fig:txy2}
\end{subfigure}
\caption{Illustration of velocity composition for multimodality tracking in different NLoS scenarios.}
\label{fig:nlos}
\vspace{-4mm}
\end{figure}

\subsection{Multimodal Tracking in the NLoS Scenario}\label{sec:eval_los}
With the advancement of smart home technology, it has become quite common for households to possess a limited number of routers and multiple smart devices, inevitably resulting in NLoS conditions between routers and smart devices.
A typical scenario involves the router being located in the living room, while multiple smart devices are dispersed in bedrooms and other areas.
Since the smart devices should communicate with the router instead of the other devices directly, we can observe many NLoS paths.

\subsubsection{Wi-Fi DPLCR}
The basic idea to realize NLoS tracking is to eliminate the shared path of different Wi-Fi links.
As illustrated in Fig.~\ref{fig:txy1},
the transmitted Wi-Fi signals will first arrive at the human and then two receivers.
Although it is different to predict the path from the transmitter to the human due to the complex NLoS reflection, we can eliminate such a path because different receivers share the same TX-User path \cite{xu2023hypertracking}.

Mathematically, we have the following equation
\begin{equation}
 \left \{
\begin{aligned}
    P_{R_{x1}} = P(T_x,U)+P(U,R_{x1}), \\
    P_{R_{x2}} = P(T_x,U)+P(U,R_{x2}),
\end{aligned}
 \right.
\label{eq:nlosboth}
\end{equation}
where $P_{R_{x1}}$ and $P_{R_{x2}}$ denote the path length for the two receivers.
$P(T_x,U)$ denotes the path from $T_x$ to $U$, $P(U,R_{x1})$ denotes the path from $U$ to $R_{x1}$, and $P(U,R_{x2})$ denotes the path from $U$ to $R_{x2}$.

Next, we can use the above two equations to eliminate the shared path components as follows:
\begin{equation}
P_{R_{x1}} - P_{R_{x2}} 
= P(U,R_{x1}) - P(U,R_{x2}).
\label{eq:hyperlos}
\end{equation}

After we eliminate the shared path in Eq.~\ref{eq:hyperlos}, we can conclude the signal relationship as the Hyperbolic Zone in Fig.~\ref{fig:txy2}.
Specifically, we have
\begin{equation}
    abs(|U_nR_{x1}| - |U_nR_{x2}| - |R_{x1}R_{x2}|) = n\frac{\lambda}{2},
\end{equation}
where $U_n$ is located at the $n^{th}$ hyperbola.

\begin{equation}
r^{(i)}(t)  = \alpha_x^{(i)} v_x(t) + \alpha_y^{(i)} v_y(t) ,
\end{equation}
where 
\begin{equation}
\left\{\begin{aligned}
&\alpha_x^{(i)} = \frac{p_{x} - p_{x_t^{(i)}}}{\|\boldsymbol{p} - \boldsymbol{p_t^{(i)}}\|} - \frac{p_{x} - p_{x_r}}{\|\boldsymbol{p} - \boldsymbol{p_r^{(i)}}\|},\\
& \alpha_y^{(i)} = \frac{p_{y} - p_{y_t^{(i)}}}{\|\boldsymbol{p} - \boldsymbol{p_t^{(i)}}\|} -\frac{p_{y} - p_{y_r}}{\|\boldsymbol{p} - \boldsymbol{p_r^{(i)}}\|}.
\end{aligned}
\right.
\label{eq:16}
\end{equation}
Accordingly, in the NLoS scenario, the user's velocity can be written as
\begin{equation}
\vec{v}(t) = (\mathcal{L}^T \mathcal{L}_{nlos})^{-1}\mathcal{L}^T \vec{r}(t).
\end{equation}

\subsubsection{Fusion Model in the NLoS Scenario}
Accordingly, the fusion model should satisfy the following characteristics:
i) the physical relationship between speed and position;
ii) the relative movement satisfies the DPLCR derived from the Wi-Fi signals;
iii) the absolute position being located within the acoustic hyperbolic zone.
Accordingly, human tracking can be summarized as the following optimization equation
\begin{equation}
\left\{
\begin{aligned}
& \hat{\vec{p}}(t) = \arg \min_{\vec{p}(t)} k_1 E_1 + k_2 E_2, \\
& E_1 = \|\vec{v}(t) - (\mathcal{L}_{nlos}^T \mathcal{L}_{nlos})^{-1}\mathcal{L}_{nlos}^T \vec{r}(t)\|, \\
& E_2 = \|\|\vec{p}(t) - \vec{p}_{s1}\|  - \|\vec{p}(t) - \vec{p}_{s2}\| - t_d(t)c_s\|, \\
& \vec{p}(t + \Delta t) = \vec{p}(t) + \vec{v}(t) \Delta t,
\end{aligned}
\right.\label{eq:optifun}
\end{equation}
where $\vec{p}_{s1}$ and $\vec{p}_{s2}$ denote the position of two speakers, $k_1$ and $k_2$ are the factors with respect to different modalities.
Our objective here is to find the optimal weights, subject to the position and velocity between adjacent time samples.

\section{Tracking Mechanism}
\label{sec:track}

\begin{figure}[t]
\begin{subfigure}[t]{.32\linewidth}
\centering\includegraphics[width=\linewidth]{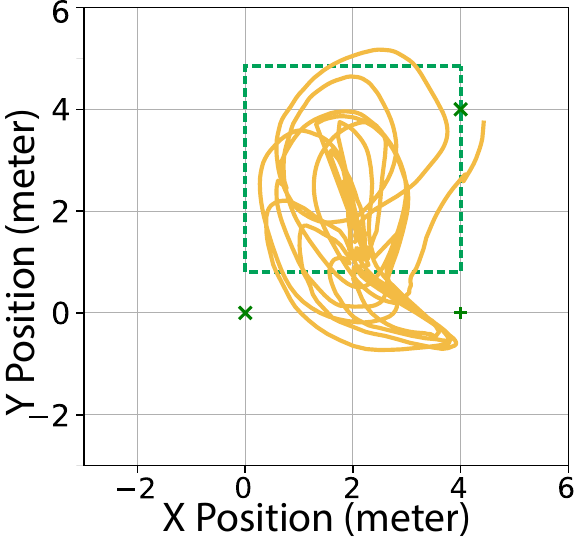}
    \caption{W/O initial position.}
    \label{fig:init_no}
\end{subfigure}
\hfill
\begin{subfigure}[t]{.32\linewidth}
\centering\includegraphics[width=\linewidth]{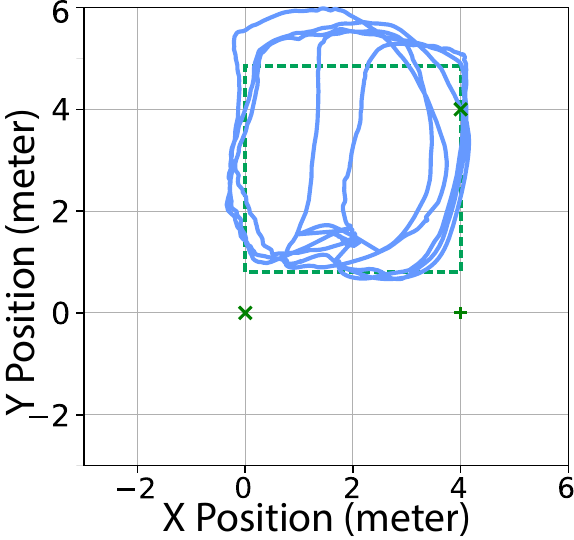}
    \caption{With given specific initial position.}
    \label{fig:init_no}
\end{subfigure}
\hfill
\begin{subfigure}[t]{.32\linewidth}
\centering\includegraphics[width=\linewidth]{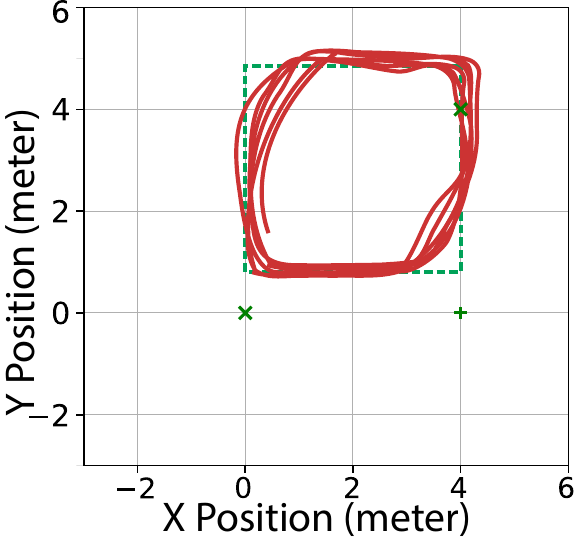}
    \caption{With reverse reconstructed initial.}
    \label{fig:init_yes}
\end{subfigure}
\caption{Impact of initial position.}
\label{fig:init}
\vspace{-4mm}
\end{figure}

\begin{figure}
    \centering
    \includegraphics[width=.96\linewidth]{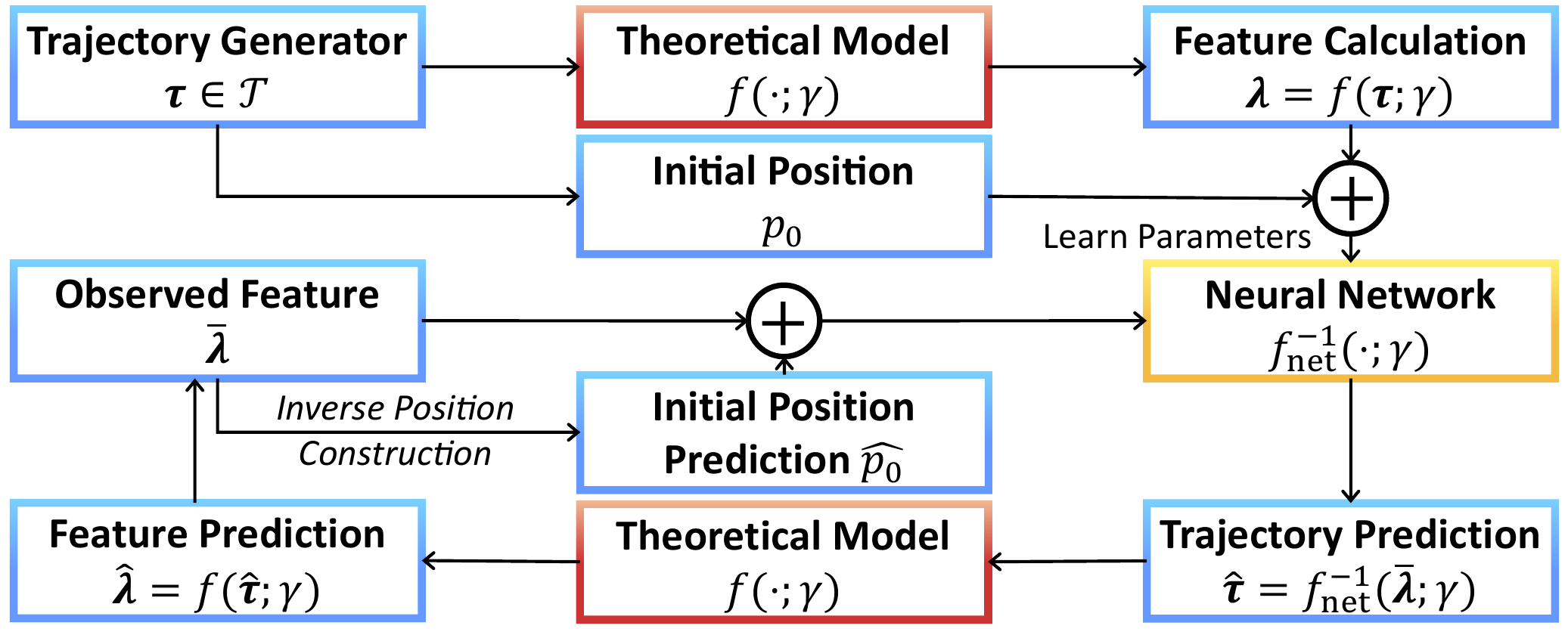}
    \caption{Structure of network.}
    \label{fig:nne}
    \vspace{-4mm}
\end{figure}
From previous sections, we have mathematically represented these features.
Building upon this, and specifically from Eq.~\ref{eq:16} and \ref{eq:optifun}, we derive two sets of optimization equations:
(1) Since we can segment the trajectory into intervals, we can express the location in the next time sample with the current position and the velocity.
Therefore, the interdependence of velocity and position complicates their direct separation;
(2) It is difficult to obtain an accurate trajectory without known initial position;
(3) Acoustic performance varies significantly at different positions, making it difficult to determine $k_1$ and $k_2$ in Eq.~\ref{eq:optifun} directly.

Though it is challenging to obtain direct solutions for the complex optimization function, we can model these conditions, including the significant acoustic impact, limited effective range, and difficulty in separating Wi-Fi features based on the simulations.
Specifically, we use these simulations to construct a solver, considering the following features  and constraints:
\begin{itemize}
\item An individual's walking pattern: we utilize the velocity and acceleration to depict the user's walking pattern;
\item Wi-Fi features: each router's PLCR is determined for each time sample based on the simulated trajectory;
\item Acoustic features: we have simultaneously designed acoustic features and confidence measures to delineate the absolute location of the user.
\end{itemize}

We aim to avoid collecting any real-world data for training.
Hence, we utilize simulated data as a means to generate both the inputs (features) and outputs (trajectories) to make up the training dataset.
%
% It is challenging to obtain trajectories from observed features due to the involved iterative processes, optimization, and the potential lack of direct solutions.
% %
% However, we discover that the forward process is more intuitive and straightforward.
%
Specifically, we can first simulate human trajectories and then derive the Wi-Fi PLCRs and Acoustic TDoF from these trajectories. Through this approach, we can completely generate training data and labels using simulations.

\begin{figure*}[t]
\centering
\begin{subfigure}[t]{.26\linewidth}
\centering\includegraphics[width=\linewidth]{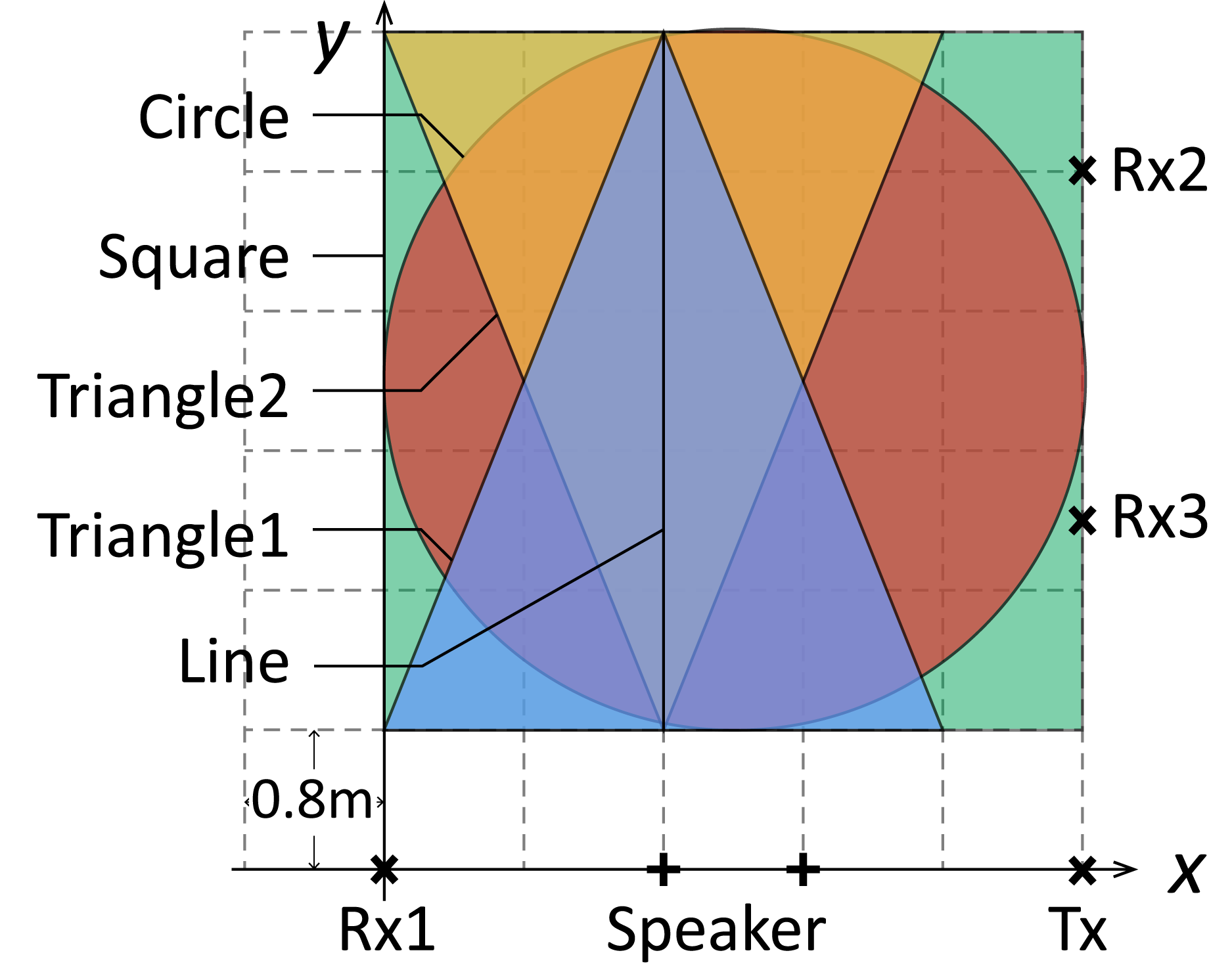}
    \caption{Sketches of LoS trajectories.}
    \label{fig:imple2}
\end{subfigure}
\hfill
\begin{subfigure}[t]{.265\linewidth}
\centering\includegraphics[width=\linewidth]{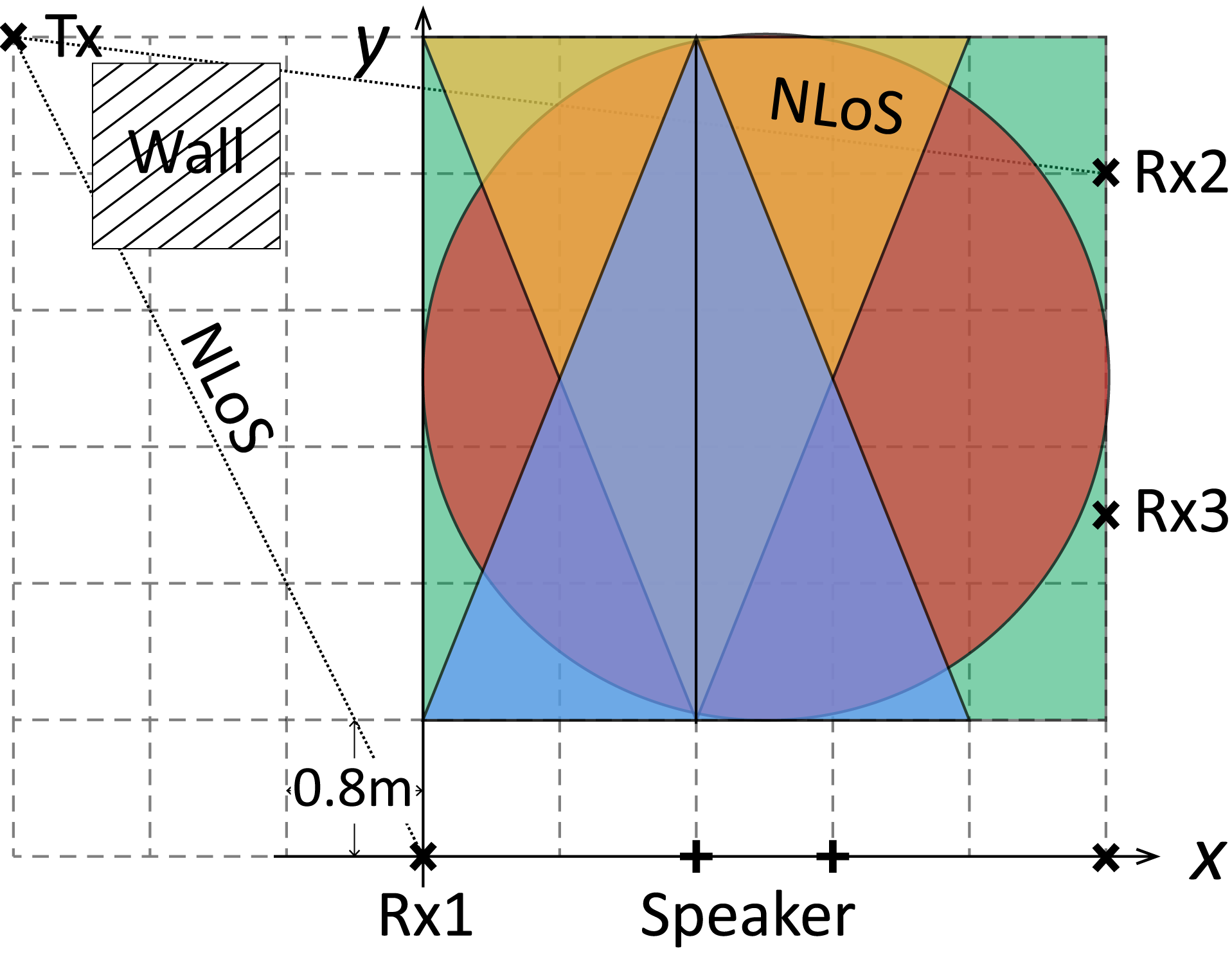}
    \caption{Sketches of NLoS trajectories.}
    \label{fig:imple3}
\end{subfigure}
\hfill
\begin{subfigure}[t]{.44\linewidth}
\centering\includegraphics[width=\linewidth]{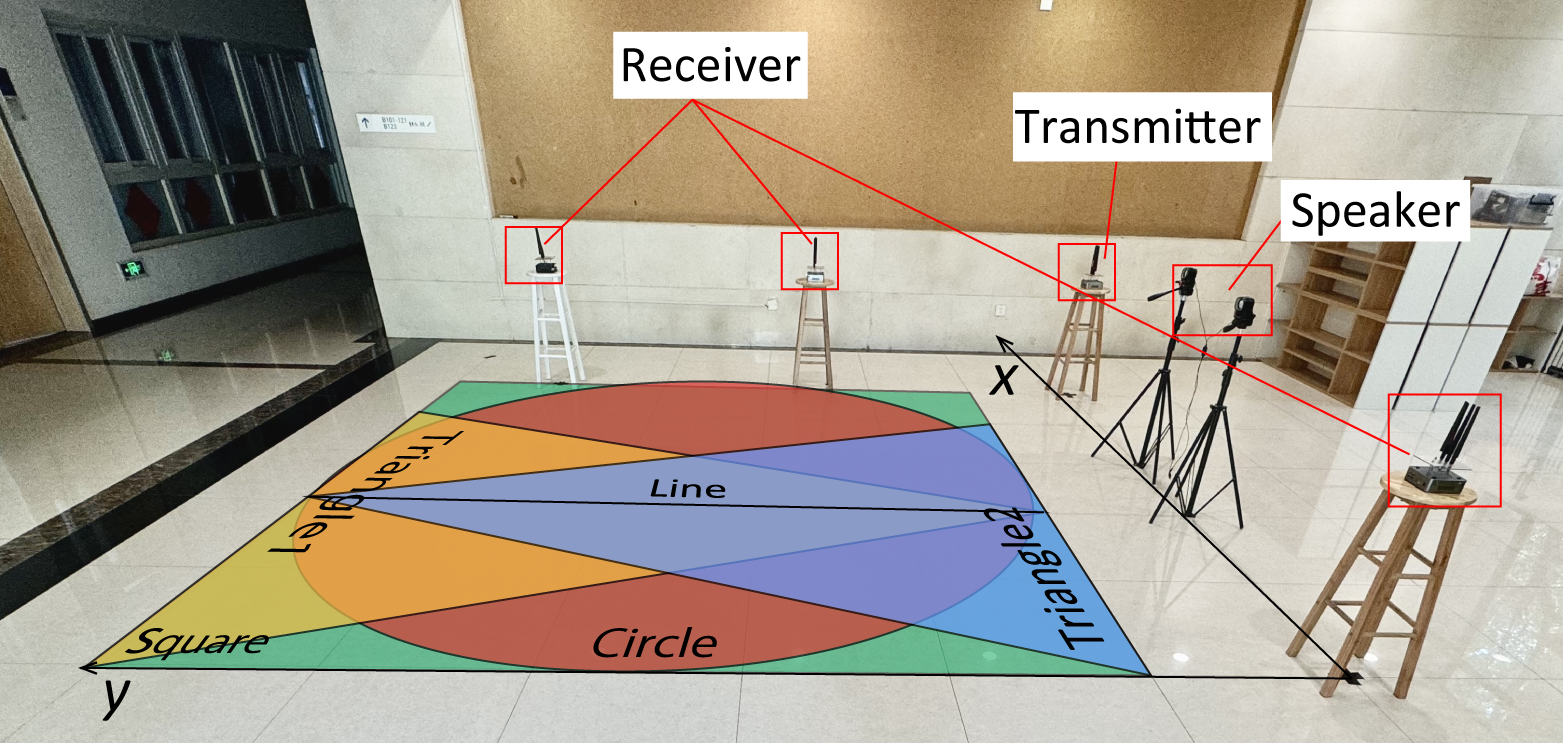}
    \caption{LoS scenario device layout of the evaluation environment.}
    \label{fig:imple1}
\end{subfigure}
\caption{Implementation and experimental setup.}
\vspace{-4mm}
\end{figure*}

It is necessary to provide the initial position for accurate prediction of the neural network.
Fig.~\ref{fig:init} demonstrates the necessity of an initial position.
In Fig.~\ref{fig:init}(a), we take 
the three-dimensional feature as the input, including two Wi-Fi PLCRs and an Acoustic TDoF.
In Fig.~\ref{fig:init}(b), we take 
the five-dimensional feature as the input, where we add the initial position’s $x$ and $y$ coordinates.
We can observe that the trajectory prediction in the Fig.~\ref{fig:init}(b) is significantly more accurate than the trajectory in the Fig.~\ref{fig:init}(a).
It is because, the acoustic feature is not always reliable, and the Wi-Fi feature exhibits relativity.
Hence, diverse positions can correspond to the same PLCR readings.

We propose a method to deduce the initial position of a system without prior knowledge by utilizing a theoretical model and a neural network. The approach is outlined as follows:
The theoretical model, denoted as \( f \), calculates physical quantities such as PLCR and TDoF from given trajectories.
Using the initial positions and the features computed from trajectories, we train a neural network \( f_{\text{net}}^{-1} \) to approximate the inverse mapping of \( f \). This network predicts trajectories based on observed features.
For a specific observed feature \(\overline{\lambda}\), the network \( f_{\text{net}}^{-1} \) is used to predict the corresponding trajectory \(\hat{\mathbf{\tau}}\). Concurrently, the initial position \(\hat{\mathbf{p}}_0\) is estimated through an inverse position construction function.
Finally, the predicted trajectory \(\hat{\mathbf{\tau}}\) is input back into the theoretical model \( f \) to compute the predicted features $\hat{\lambda}$.
This step serves to verify the accuracy of the predictions and to iteratively adjust the model parameters for enhanced prediction fidelity.
This method allows for robust inference of initial system positions and the simulation of trajectory dynamics from observed features, improving the model’s applicability in physical and simulation environments.

\begin{equation}
\left\{
\begin{aligned}
& \hat{p}_0 = \arg \min_{\gamma} L(\gamma),\\
& L(\gamma) = \| f \left(f_{\text{net}}^{-1}\left(\overline{\lambda};\gamma\right);\gamma\right) - \overline{\lambda}\|.
\end{aligned}\right.
\end{equation}
% \begin{equation}
% \left\{
% \begin{aligned}
% & \hat{\mathbf{P}} = \arg \min_{\mathbf{P}} \mathcal{L}(\mathbf{P}; f(\mathcal{T})),\\
% & \mathcal{L}(\mathbf{P}; f(\mathcal{T})) = \sum_{\mathcal{T}} \left|f(\mathcal{T}) - \hat{f}(\mathcal{T}; \mathbf{F})\right|.
% \end{aligned}\right.
% \end{equation}
By comparing predicted features with actual observations, the model iteratively refines its accuracy, ensuring improvement over time without real-world data collection.

We discovered that the reliability of acoustic signals diminishes rapidly beyond 1 meter.
To account for this, we introduce a \textit{confidence metric} based on the amplitude of received acoustic signals.
An amplitude threshold of approximately $2000$ indicates the target’s effective range within the acoustic system, making confidence a crucial input dimension.

In summary, the proposed \textit{DuTrack} model integrates Wi-Fi and acoustic signals to provide high-accuracy tracking. By incorporating initial position data, inverse position construction, and confidence metrics for acoustic features, the model achieves robust tracking performance.
\section{Evaluation}\label{sec:experiment}
\subsection{Implementation}
Our system comprises a Wi-Fi transmitter and a receiver, both equipped with Intel 5300 NICs. The Wi-Fi transmitter is a single-antenna device, while the receiver features three antennas, facilitating robust signal processing.
The Wi-Fi transmitter broadcasts packets at a rate of 1,000 packets per second, while the receiver, with its three antennas, is positioned to optimize sensing performance.
The entire system's software is developed and executed using Matlab, providing a robust platform for our experiments. 
For acoustic sensing, we use the built-in dual microphones of a smartphone, which moves with the user to capture dynamic acoustic data.
This approach leverages the phone's existing hardware, ensuring convenient and efficient data capture as the user moves through different environments.
All components are connected to a laptop featuring an Intel i7-11800H CPU and 16GB RAM, which serves as the central hub for data collection and processing. To manage packet transmission, we utilize an SSH tool compatible with the linux-802.11n protocol on Ubuntu 14.04.3. The laptop is configured to work in monitor mode on a channel with minimal interference, ensuring the integrity of the collected data.

\subsubsection{Evaluation Setup}

To thoroughly evaluate the performance of \textit{DuTrack}, we conducted experiments in two different indoor environments: LoS and NLoS scenarios. In the LoS scenario, shown in Fig.\ref{fig:imple2}, there are direct links between the transmitter and the routers. In contrast, the NLoS scenario, depicted in Fig.\ref{fig:imple3}, involves a strict wall obstruction between the transmitter and the receivers, affecting their relative positions.
The room layout, sensing area, and general signal propagation paths are illustrated in Fig.~\ref{fig:imple1}. Additionally, four trajectories are shown in the figures.

\subsubsection{Dataset}
We collect data from four trajectories at each location, as shown in Fig.~\ref{fig:imple2}. The designed trajectories include square, circular, and two types of triangular paths, covering common scenarios with both sharp and smooth turns. The dataset consists of tracing samples from $4$ individuals, each following $4$ trajectories for $8$ laps, repeated $3$ times, using $3$ different device combinations.
Therefore, the dataset contains over $1000$ samples.

To accurately evaluate the performance of \textit{DuTrack}, it is necessary to compare the predicted paths with the actual paths. Therefore, we recorded the ground truth of the user's walking trajectory to assess the location errors at each timestamp. 

\subsection{Overall Performance}

\begin{figure*}[t]
\centering
\begin{subfigure}[t]{.18\linewidth}
\centering\includegraphics[width=\linewidth]{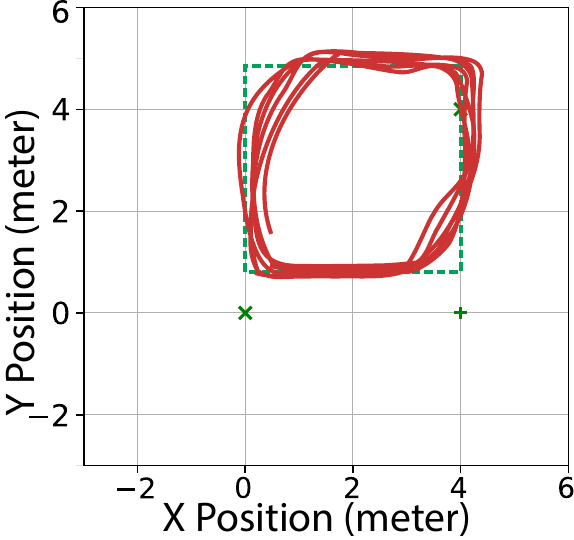}
    \caption{Square.}
\end{subfigure}
\hfill
\begin{subfigure}[t]{.18\linewidth}
\centering\includegraphics[width=\linewidth]{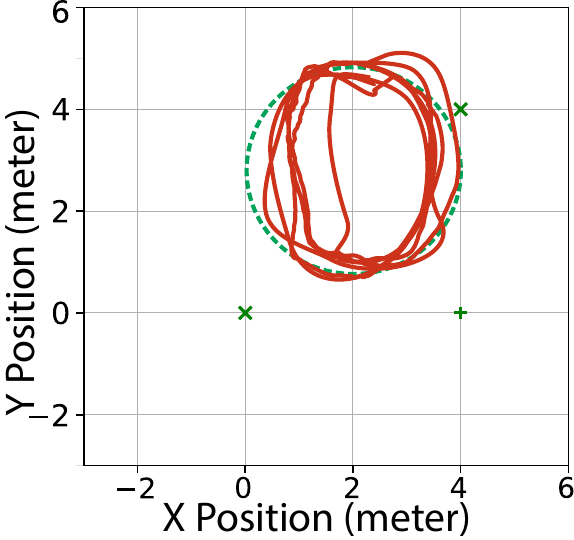}
    \caption{Circle.}
\end{subfigure}
\hfill
\begin{subfigure}[t]{.18\linewidth}
\centering\includegraphics[width=\linewidth]{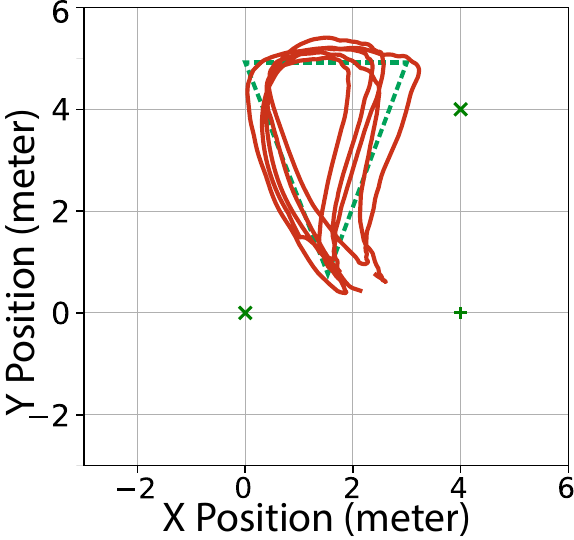}
    \caption{Triangle.}
\end{subfigure}
\hfill
\begin{subfigure}[t]{.18\linewidth}
\centering\includegraphics[width=\linewidth]{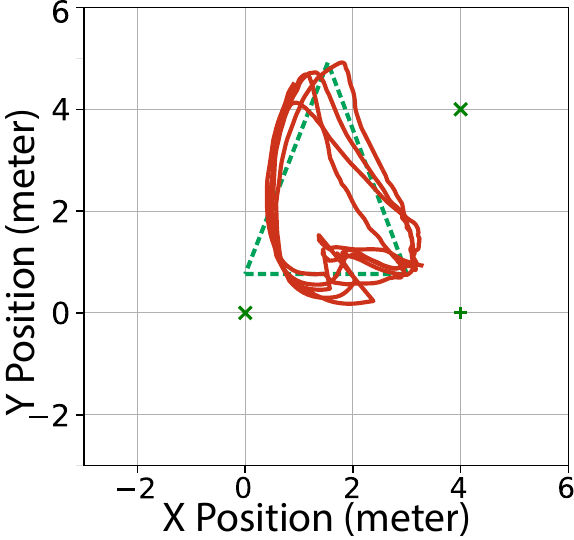}
    \caption{Triangle (inverted).}
\end{subfigure}
\hfill
\begin{subfigure}[t]{.18\linewidth}
\centering\includegraphics[width=\linewidth]{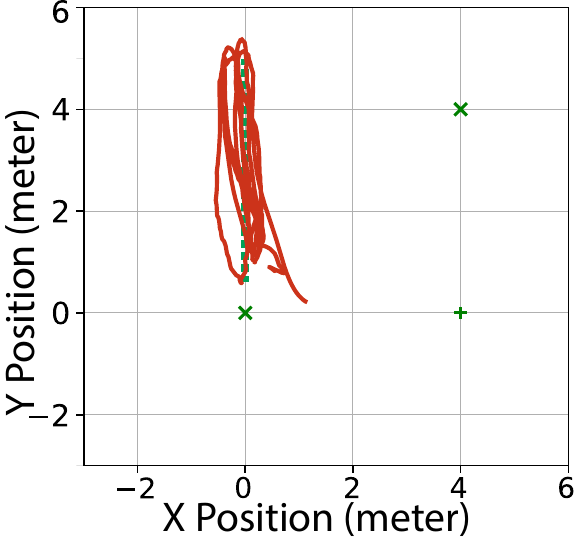}
    \caption{Line.}
\end{subfigure}
\caption{Examples of tracking results of \textit{DuTrack}.}
\label{fig:trace1}
\end{figure*}

\begin{figure*}[t]
\centering
\begin{subfigure}[t]{.18\linewidth}
\centering\includegraphics[width=\linewidth]{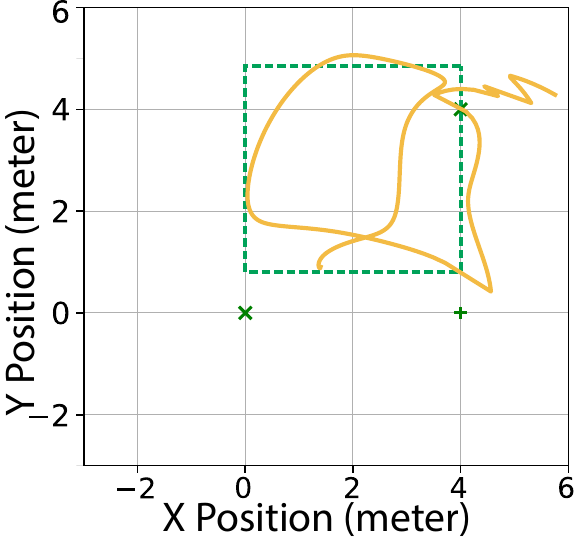}
    \caption{Square.}
\end{subfigure}
\hfill
\begin{subfigure}[t]{.18\linewidth}
\centering\includegraphics[width=\linewidth]{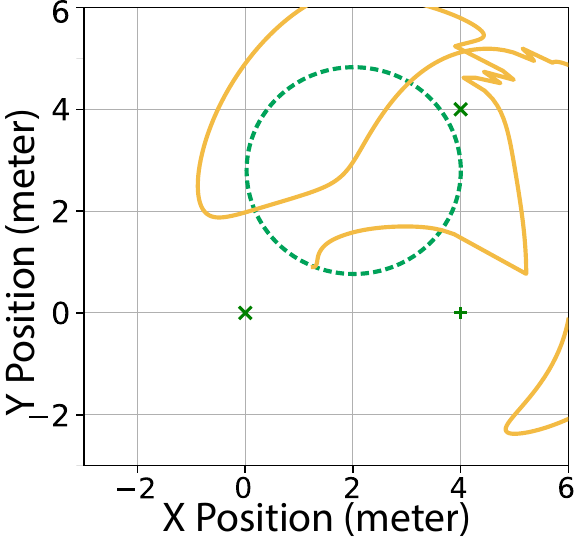}
    \caption{Circle.}
\end{subfigure}
\hfill
\begin{subfigure}[t]{.18\linewidth}
\centering\includegraphics[width=\linewidth]{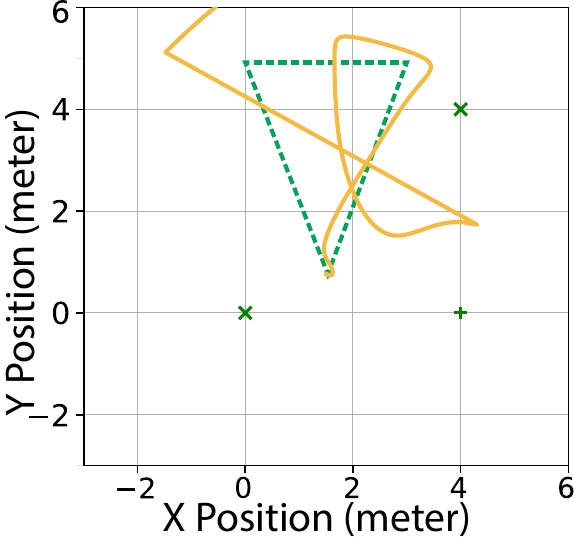}
    \caption{Triangle.}
\end{subfigure}
\hfill
\begin{subfigure}[t]{.18\linewidth}
\centering\includegraphics[width=\linewidth]{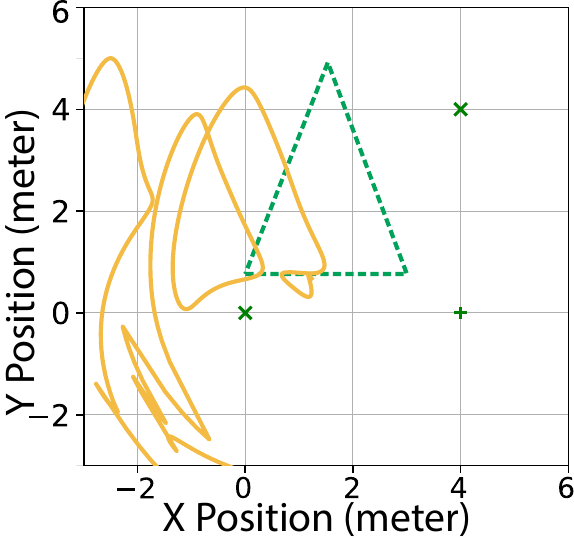}
    \caption{Triangle (inverted).}
\end{subfigure}
\hfill
\begin{subfigure}[t]{.18\linewidth}
\centering\includegraphics[width=\linewidth]{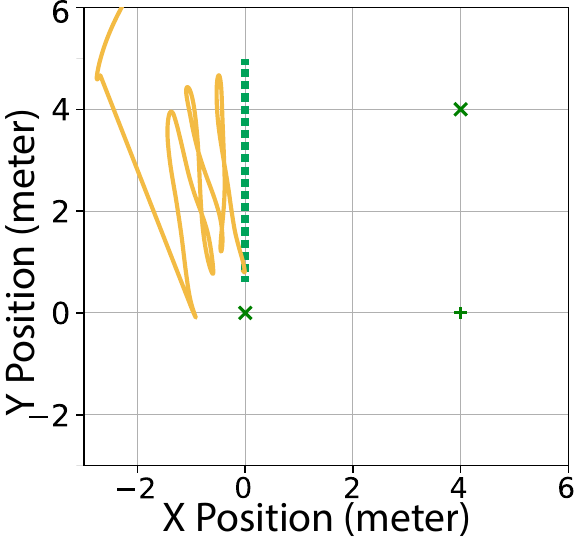}
    \caption{Line.}
\end{subfigure}
\caption{Examples of tracking results of model-driven methods.}
\label{fig:trace2}
\end{figure*}

\begin{figure*}[t]
\centering
\begin{subfigure}[t]{.18\linewidth}
\centering\includegraphics[width=\linewidth]{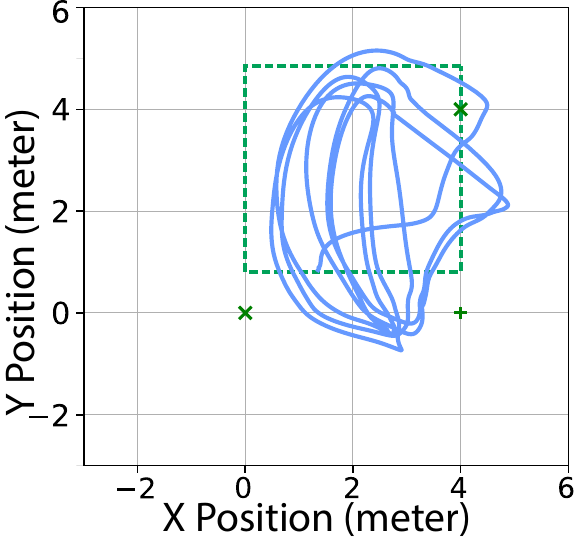}
    \caption{Square.}
\end{subfigure}
\hfill
\begin{subfigure}[t]{.18\linewidth}
\centering\includegraphics[width=\linewidth]{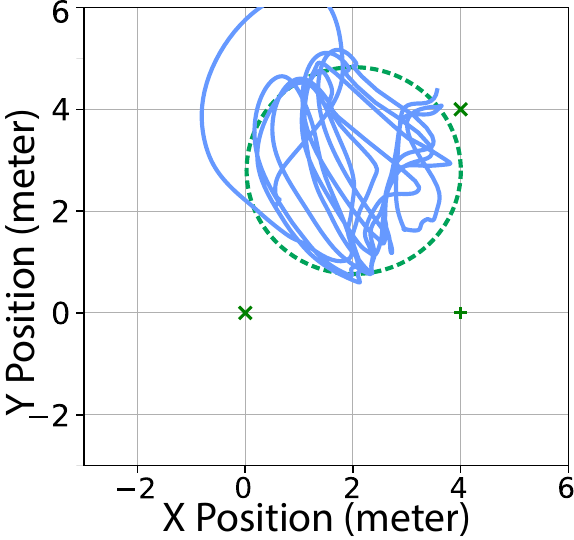}
    \caption{Circle.}
\end{subfigure}
\hfill
\begin{subfigure}[t]{.18\linewidth}
\centering\includegraphics[width=\linewidth]{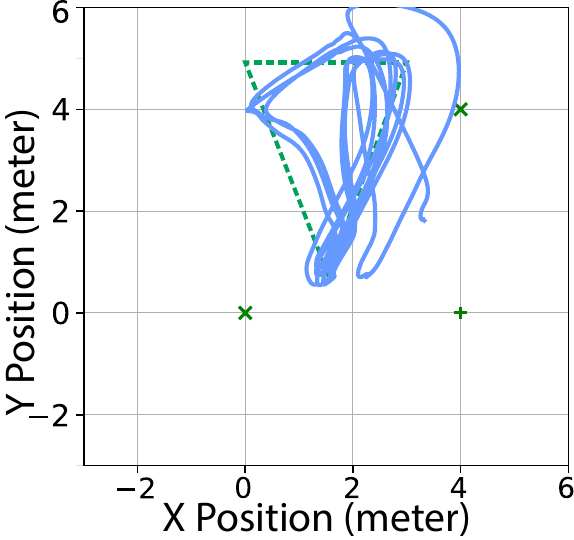}
    \caption{Triangle.}
\end{subfigure}
\hfill
\begin{subfigure}[t]{.18\linewidth}
\centering\includegraphics[width=\linewidth]{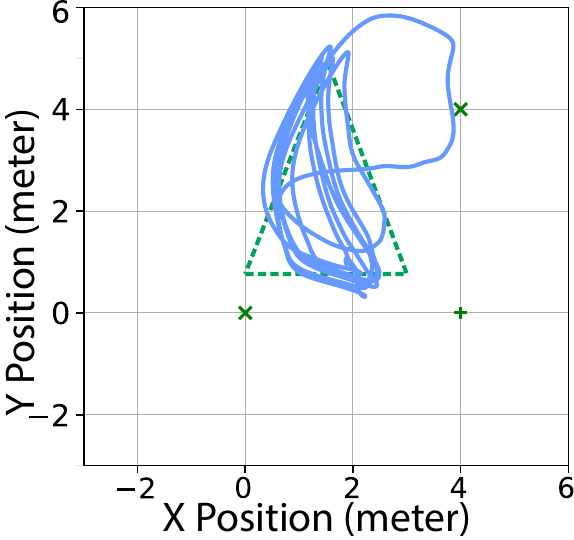}
    \caption{Triangle (inverted).}
\end{subfigure}
\hfill
\begin{subfigure}[t]{.18\linewidth}
\centering\includegraphics[width=\linewidth]{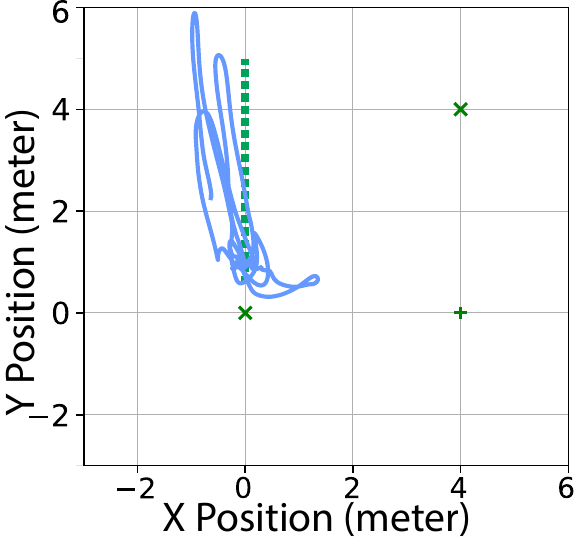}
    \caption{Line.}
\end{subfigure}
\caption{Examples of tracking results of data-driven methods (NNE-Tracking).}
\label{fig:trace3}
\vspace{-4mm}
\end{figure*}

\begin{figure}[t]
\centering
\begin{subfigure}[t]{.48\linewidth}
\centering\includegraphics[width=\linewidth]{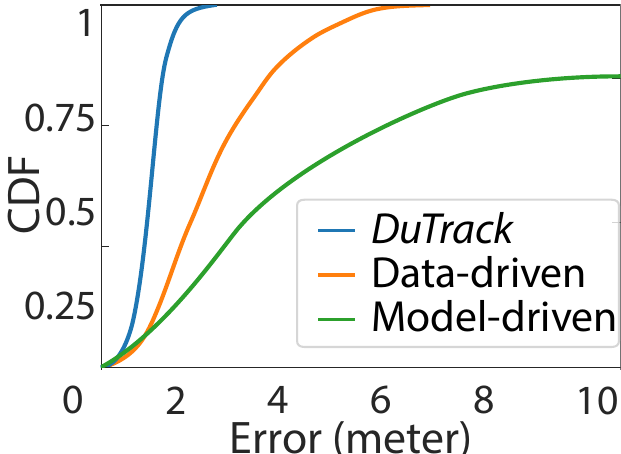}
    \caption{LoS scenario .}
    \label{fig:cdf1}
\end{subfigure}
\hfill
\begin{subfigure}[t]{.48\linewidth}
\centering\includegraphics[width=\linewidth]{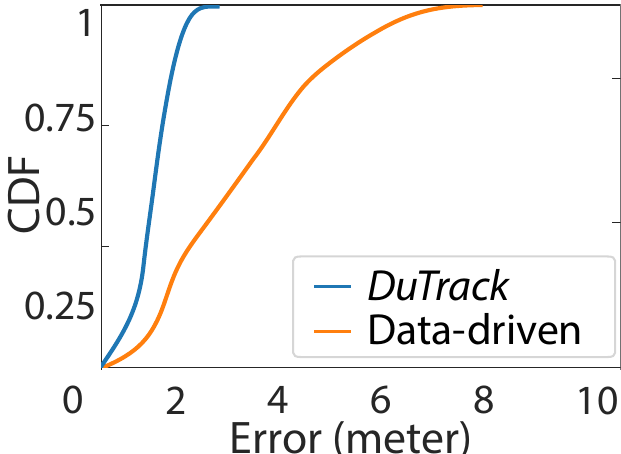}
    \caption{NLoS scenario.}
    \label{fig:cdf1}
\end{subfigure}
\caption{Overall localization accuracy.}
\label{fig:cdf}
\vspace{-4mm}
\end{figure}

\begin{figure}[t]
\begin{subfigure}[t]{.32\linewidth}
\centering\includegraphics[width=\linewidth]{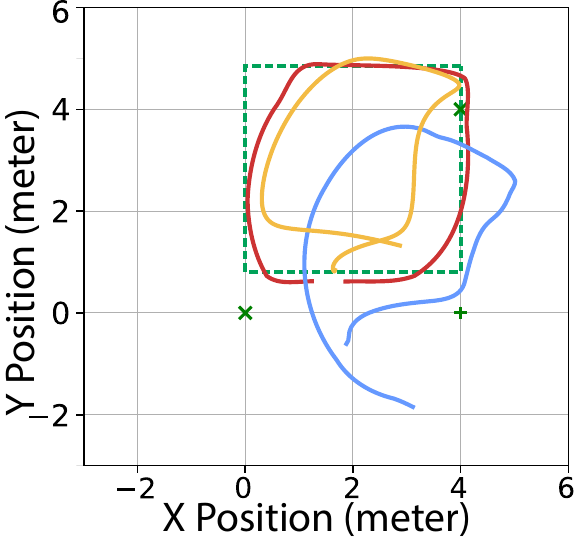}
    \caption{The first lap.}
\end{subfigure}
\hfill
\begin{subfigure}[t]{.32\linewidth}
\centering\includegraphics[width=\linewidth]{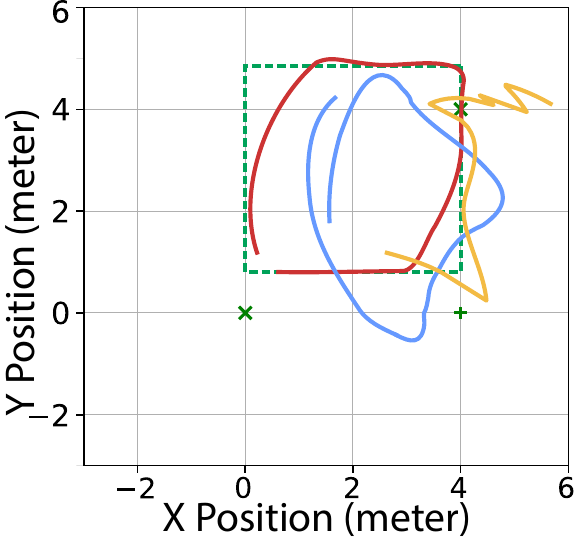}
    \caption{The third lap.}
\end{subfigure}
\hfill
\begin{subfigure}[t]{.32\linewidth}
\centering\includegraphics[width=\linewidth]{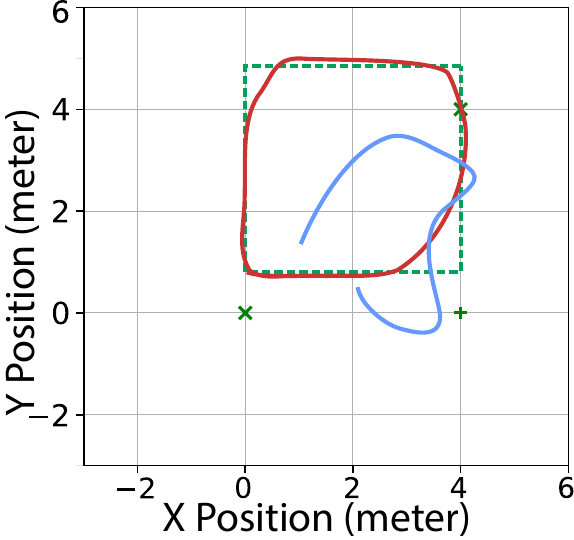}
    \caption{The sixth lap.}
    \label{fig:cdf3}
\end{subfigure}
\caption{Comparison with data- and model-driven methods.}
\label{fig:lap}
\vspace{-4mm}
\end{figure}

Taking all factors into consideration, \textit{DuTrack} achieves an average localization error of $0.78$ meters.
Examples of tracking results are shown in Fig.~\ref{fig:trace1}. 
We present five types of trajectories (i.e., square, circle, triangle (both upright and inverted), and line). Each trajectory involves two minutes of repetitive movements, with variations in the number of laps depending on the path length. For comparison, we plotted the same set of data using both the data-driven model and the model-driven approach, as shown in Fig.~\ref{fig:trace2} and Fig.~\ref{fig:trace3}, respectively.
Fig.~\ref{fig:cdf} shows the Cumulative Distribution Function (CDF) graphs considering each method in both LoS and NLoS scenarios. Since data-driven methods only consider LoS scenarios, we compare the CDF of \textit{DuTrack} and model-driven methods in the NLoS scenario. We observe that \textit{DuTrack} achieves the highest accuracy and maintains its performance in NLoS scenarios.

\subsection{Comparative Analysis}
\begin{figure*}[t]
\centering
\begin{subfigure}[t]{.24\linewidth}
\centering\includegraphics[width=\linewidth]{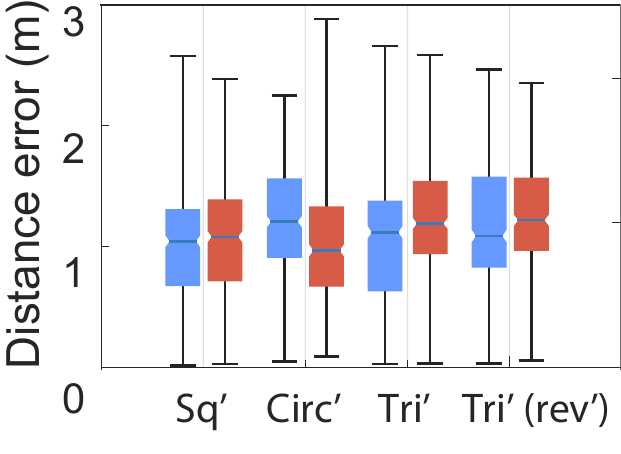}
    \caption{Impact of LoS and NLoS.}
    \label{fig:los}
\end{subfigure}
\hfill
\begin{subfigure}[t]{.24\linewidth}
\centering\includegraphics[width=\linewidth]{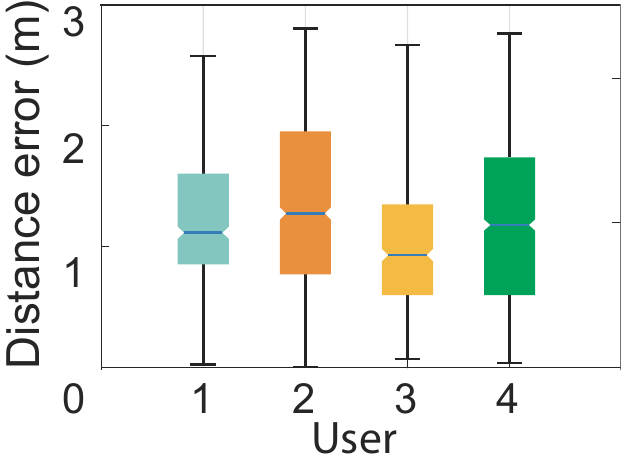}
    \caption{Impact of individuals.}
    \label{fig:user}
\end{subfigure}
\hfill
\begin{subfigure}[t]{.24\linewidth}
\centering\includegraphics[width=\linewidth]{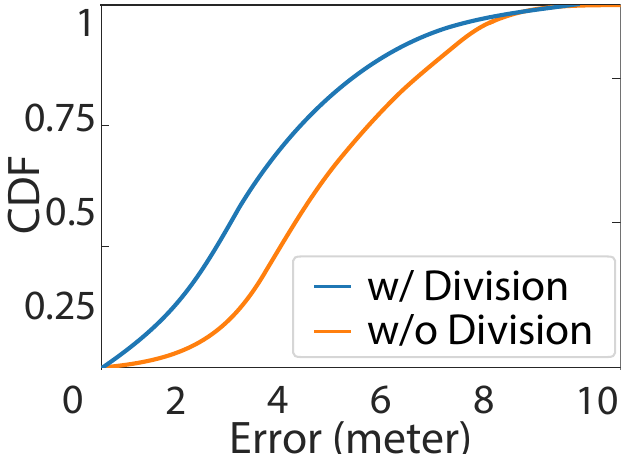}
    \caption{Benefit of data division.}
    \label{fig:datadivision}
\end{subfigure}
\hfill
\begin{subfigure}[t]{.24\linewidth}
\centering\includegraphics[width=\linewidth]{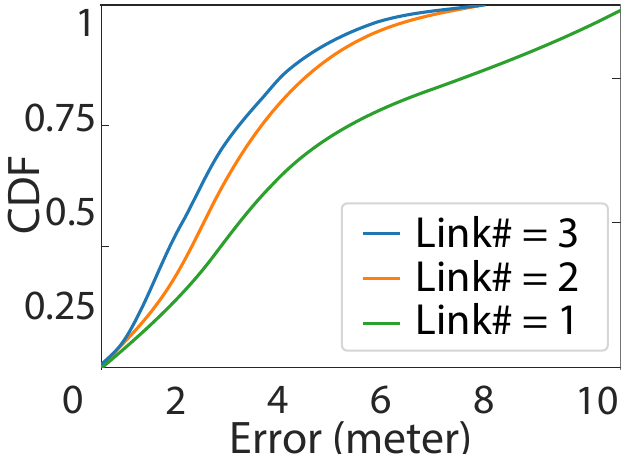}
    \caption{Impact of number of links.}
    \label{fig:link}
\end{subfigure}
\caption{Parameter analysis.}
\label{fig:trace}
\vspace{-3mm}
\end{figure*}

\subsubsection{Comparison of Model-Driven Method}
Some existing works propose purely theoretical methods to track individuals~\cite{qian2018widar2}.
However, they exhibit several limitations:
i) They require extensive traversals and computations to achieve optimal estimation. This substantially increases both the computational cost per prediction and the overall time required;
ii) Model-based methods often overlook the inherent physical properties of human motion, leading to theoretical strategies that are not consistently robust.

Conversely, a data-driven tracking framework utilizing neural networks demonstrates a significantly lower execution time post-training. 
The training data pushes the boundaries of human motion capabilities, such as motion inertia, which the neural network can learn through simulated parameters.
As depicted in Fig.~\ref{fig:trace3}, the accuracy of the predicted results sharply decreases with the increasing number of laps. For long-term tracking, the methods can only ensure accurate data for approximately one lap.
This prompts the consideration of developing a data-driven tracking method that learns the characteristics of the hyperbolic tracking model, thereby estimating accurate user trajectories in a cost-effective manner.

\subsubsection{Comparison of Data-Driven Methods}
We compare \textit{DuTrack} with an alternative state-of-the-art methodology, NNE~\cite{tong2024nne}. Specifically, NNE uses three PLCR features as learning inputs and adopts the inverted reconstruction LSTM model. Fig.~\ref{fig:cdf} shows the system performance of the two approaches; \textit{DuTrack} achieves better performance than NNE. 
This is because \textit{DuTrack} leverages calibration with acoustic features to obtain definitive information for tracking. 
This also demonstrates the importance of absolute information in tracking, especially for long-term tracking.

We further analyze a few specific laps to better demonstrate the effectiveness of the three methods, and use two criteria to evaluate their effectiveness: (1) the ability to depict the shape, for example, all four sides of a square, and (2) the error compared to the ground truth.
Initially, we plot the complete trajectories for all laps and then select specific laps for detailed analysis. 
We present the predicted results for the first, third, and fifth laps of the square trajectory for each method in Fig.~\ref{fig:lap}. 
During the first lap, all three methods are able to describe the square shape, with average errors of 0.24, 1.18, and 3.12 meters, respectively. However, by the third lap, the other two methods could no longer accurately depict all four sides of the square. By the fifth lap, the predicted trajectory of the data-driven method has exceeded the plotted area.

\subsection{Parameter Analysis}

\subsubsection{LoS and NLoS}
As shown in Fig.~\ref{fig:imple3}, to verify the performance of \textit{DuTrack} in NLoS scenarios, we move the transmitter behind a wall, so that the direct path between the transmitter and both receivers are blocked.
Results are shown in Fig.~\ref{fig:los}. We use box plots to show the prediction errors for each specific trajectory, where the blue boxes represent the results for the LoS scenario, and the red boxes represent the results for the NLoS scenario. 
We observe that the prediction errors exhibit strong consistency, and the impact of LoS or NLoS scenarios on the results is minimal.

\subsubsection{User Variability}
Data collected from different individuals may exhibit discrepancies due to their unique characteristics. However, \textit{DuTrack} does not require data collected for training, ensuring that our model remains unbiased. To evaluate the performance of \textit{DuTrack} across different users, we tested the model with data from four individuals. The performance across these different users is shown in Fig.~\ref{fig:user}.

\subsubsection{Data Division Methods}
One advantage of LSTM is its ability to handle input data of varying lengths. 
Nonetheless, varying data lengths can impact the model's performance. 
To address this, we propose a method for segmenting long data into smaller, manageable chunks while preserving performance integrity. 
Due to the limited recognition range of acoustic signals and quality degradation at longer distances, it becomes challenging to extract useful information. 
Our method involves segmenting the data such that each chunk contains at least one period of high-confidence signals, ensuring each segment retains valuable acoustic features for effective calibration.
Figure~\ref{fig:datadivision} compares the outcomes of using confidence-based data segmentation versus the original undivided data. The results indicate that segmentation improves performance.

\subsubsection{Link Numbers}
We further compare different number of Wi-Fi routers and demonstrate the system performance in Fig.~\ref{fig:link}.
We aim to verify whether the number of input features, i.e., PLCRs, can be reduced while still achieving effective tracking. Specifically, we set the number of PLCR features to 1, 2, and 3, respectively, to evaluate their impact on performance.
Each PLCR necessitates a pair of Wi-Fi transceivers. Therefore, for 1, 2, and 3 links, a single transmitter paired with 1, 2, and 3 receivers respectively is required.
Using two and three links has average error of 0.78 and 0.75 meters. However, when the number of links is reduced to one, the performance significantly deteriorates. This is due to the limited sensing range of a single link, resulting in inaccurate predictions for points beyond its range.

\section{Conclusion}\label{sec:conclude}

We introduced \textit{DuTrack}, an innovative tracking system that combines acoustic and Wi-Fi signals.
This hybrid approach addresses the limitations of existing methods, providing accurate long-term indoor tracking through a specially designed inverse construction network and an adaptive confidence indicator.
Simulations and experiments show that \textit{DuTrack} achieves superior accuracy, with an average error of 0.78 meters. It integrates seamlessly with home audio and router systems, making it a cost-effective and accessible solution.
Implemented on commodity devices, \textit{DuTrack} offers a reliable, robust, and economical solution for long-term human tracking.
We achieved an $89.37\%$ reduction in median tracking error compared to model-based methods and a $65.02\%$ reduction compared to data-driven methods.
With its high accuracy in long-term applications, \textit{DuTrack} marks a critical step forward in the evolution of tracking systems.

\balance

% Switch to single-column mode and center the References title
\twocolumn[
  \begin{@twocolumnfalse}
    \centering
    \section*{References}
    \vspace{1em} % Adjust this space if necessary
  \end{@twocolumnfalse}
]

\begingroup
\renewcommand{\section}[2]{}%
\bibliography{references}
\endgroup
\bibliographystyle{ieeetr}

\end{document}